\documentclass[fleqn,usenatbib]{mnras}
\usepackage{newtxtext,newtxmath}

\usepackage[T1]{fontenc}

\DeclareRobustCommand{\VAN}[3]{#2}
\let\VANthebibliography\thebibliography
\def\thebibliography{\DeclareRobustCommand{\VAN}[3]{##3}\VANthebibliography}

\usepackage{color}
\usepackage{booktabs}
\usepackage{enumitem}
\usepackage{hyperref}
\usepackage{verbatim}
\usepackage{amsmath,amstext,mathrsfs}
\usepackage[all]{hypcap} 
\usepackage{afterpage}    
\usepackage{marginnote}
\usepackage{makecell}
\usepackage{subfigure}
\usepackage{xfrac}

\usepackage{ifxetex,ifluatex}
\usepackage{fixltx2e} 
\usepackage{multirow}

\DeclareFontFamily{OMS}{oasy}{\skewchar\font48 }
\DeclareFontShape{OMS}{oasy}{m}{n}{%
         <-5.5> oasy5     <5.5-6.5> oasy6
      <6.5-7.5> oasy7     <7.5-8.5> oasy8
      <8.5-9.5> oasy9     <9.5->  oasy10
      }{}
\DeclareFontShape{OMS}{oasy}{b}{n}{%
       <-6> oabsy5
      <6-8> oabsy7
      <8->  oabsy10
      }{}
\DeclareSymbolFont{oasy}{OMS}{oasy}{m}{n}
\SetSymbolFont{oasy}{bold}{OMS}{oasy}{b}{n}

\DeclareMathSymbol{\smallleftarrow}     {\mathrel}{oasy}{"20}
\DeclareMathSymbol{\smallrightarrow}    {\mathrel}{oasy}{"21}
\DeclareMathSymbol{\smallleftrightarrow}{\mathrel}{oasy}{"24}

\newcommand*{\torefereeone}{ }
\newcommand*{\torefereetwo}{ }

\graphicspath{{./}{fig/}}

\title[UNM Instability]{How the existence of unstable neutral media restricts the aspect ratio of cold neutral media?}
\author[Ho et.al]{
Ka Wai Ho,$^{1}$\thanks{kho33@wisc.edu}
Ka Ho Yuen,$^{1,2}$\thanks{kyuen@lanl.gov (Oppenheimer Fellow)}
Alex Lazarian$^{1,3}$\thanks{alazarian@facstaff.astro.wisc.edu}
\\
$^{1}$Department of Astronomy, University of Wisconsin-Madison, USA\\
$^{2}$Theoretical Division, Los Alamos National Laboratory, Los Alamos, NM 87545, USA\\
$^{3}$Centro de Investigación en Astronomía, Universidad Bernardo O’Higgins, Santiago, General Gana 1760, 8370993, Chile
}

\date{Accepted XXX. Received YYY; in original form ZZZ}

\begin{document}
\label{firstpage}
\pagerange{\pageref{firstpage}--\pageref{lastpage}}
\maketitle

\date{Accepted XXX. Received YYY; in original form ZZZ}

\pubyear{2021}


\begin{abstract}
The ubiquity of very thin and lengthy cold neutral {\torefereeone medium} (CNM) has been reported by multiple authors in the HI community. Yet, the reason of how the CNM can be so long and lengthy is still in debate. In this paper, we recognize a new type of instability due to the attractive nature of the pressure force in the unstable phase. We provide a new estimation of the average CNM filament aspect ratio with the consideration of force balances at the phase boundary, which is roughly 5-20 in common CNM environment. We show that most of the cold filaments are less filamentary than what usually predicted via MHD turbulence theory or inferred from observations: The average length of CNM filament is roughly 1/2 of that in isothermal MHD turbulence with similar turbulence conditions. This suggests that the "cold filaments" that {\torefereeone are} identified in observations might not be in pressure equilibrium or generated via other mechanisms. 
\end{abstract}





\section{Introduction}


The atomic hydrogen cold neutral {\torefereeone medium} (HI-CNM) is one of the most popular and important astrophysical objects that came to the researchers’ eyes in the last decade. {\torefereeone The multiphase HI media formed under the presence of heating and cooling term and naturally three phases, namely, warm, unstable and cold neutral media formed (See, e.g. Fields 1965.} Emission line observations like HP4PI \citep{KH15}, GALFA \citep{Peek18}, THOR-HI \citep{Beuther16,Wang20}, FAST (e.g. \citealt{Li21}) have advanced our understanding of CNM including its spatial distribution \citep{Kalberla16,Kalberla18}, relation to magnetic field \citep{HT03,Clark15} and its connection to underlying molecular phase \citep{Kritsuk17}. CNM is ubiquitous in interstellar media, spanning from high latitude \citep{Clark15,Kalberla17} to the galactic plane \citep{Soler20}, and being highly filamentary and linear along the magnetic field directions \citep{Clark15,YL17a}. The unique feature of CNM makes it to be one of the most important B-field probes in modern astronomy, with vast applications from cosmology \citep{CH19} to molecular cloud studies (e.g. \cite{LY18a}). Notice that observationally detecting the CNM requires the absorption line studies \citep{2014ApJ...781L..41M,2020ApJS..250....9S}, as the emission lines are shown to be capable in tracing only a part of the CNM emissions \citep{SF20}


The cold neutral media has a generally large aspect ratio along B-field, which is suggested by a number of authors observationally \citep{Clark15,Soler21}. The aspect ratio of CNM filaments is in average 60 (visual inspection of \citealt{Clark14}) and can be as large as 200-400 (Chris Mckee, private communication). However, the origin of why CNM filaments are having such a large aspect ratio is still on debate. Arguments based on the theory of MHD turbulence \citep{Xu19} suggested that the multiphase media has a continuous MHD cascade from warm to cold phase, thus the aspect ratio of overdensity filament is given by the \cite{GS95} (GS95) scaling:
\begin{equation}
    \frac{l_\parallel}{l_\perp} \sim M_A^{-4/3} (\frac{L_{inj}}{l_\perp})^{1/3}
\end{equation}
where $l_{\parallel,\perp}$ are the statistical length and width of the filaments, $M_A$ is the Alfvenic Mach number and $L_{inj}$ is the injection scale.However, even with the most extreme measure of Alfvenic Mach number $M_A$ for CNM (e.g. $M_A=0.2, L_{inj}=100, l_\perp = 0.1pc$, see \cite{Kalberla18} ), the aspect ratio is never approaching 100. Suggestions on the dynamo effect like \cite{Kalberla21} in amplifying the length of cold filaments are recently proposed, but there are no quantitative or numerical estimates on why cold filaments has to be that lengthy.


One of the most important physical quantity that the previous analysis have omitted is the unstable neutral media (UNM). Both numerically \citep{Kritsuk17} and observationally \citep{Kalberla18,Murray18} showed that UNM occupies a large fraction both in terms of mass and volume {\torefereeone ($\sim 40\%$ for both, see, e.g. \citealt{Kalberla18}, at least $20\%$ for mass fraction, \citealt{Murray18})}. However, the physical understanding on UNM is extremely inadequate since the existence of UNM is only being considered as a transient stage between the more stable two phases. In fact, UNM carries a special physical property that restricts the length of cold filaments in observations: The pressure force of UNM is attractive rather than repulsive in the two other phases, meaning that the force field of UNM is similar to that of the gravity term, contracting to higher density region (Fig \ref{fig:illustration}). Given the large mass and volume fraction of UNM as in observation and the specialities of the UNM dynamics, we must rethink the role of UNM during the formation of the CNM filaments. Most importantly, since the UNM's force term is similar to that of the gravity term, there is a natural instability that we shall name "UNM instability" in the later section where it restricts the formation of lengthy cold neutral filaments or the filament will expend perpendicularly. 


In this paper, we shall explore both theoretically and numerically on how the "UNM instability" impacts the formation of long cold filaments in the sky. \S \ref{sec:theory} discusses the physical picture and the theoretical formulation of the "UNM instability". \S \ref{sec:num} proves that the critical balance between the UNM and CNM at the phase boundary will limits the length of the CNM filament with our numerical simulations. In \S \ref{sec:dis} we discuss the impact of our result and in \S \ref{sec:con} we conclude our paper.


\section{Theoretical Treatment}
\label{sec:theory}

\subsection{UNM will collapse indefinitely if no formation of CNM happens}

\begin{figure}
\includegraphics[width=0.5\textwidth]{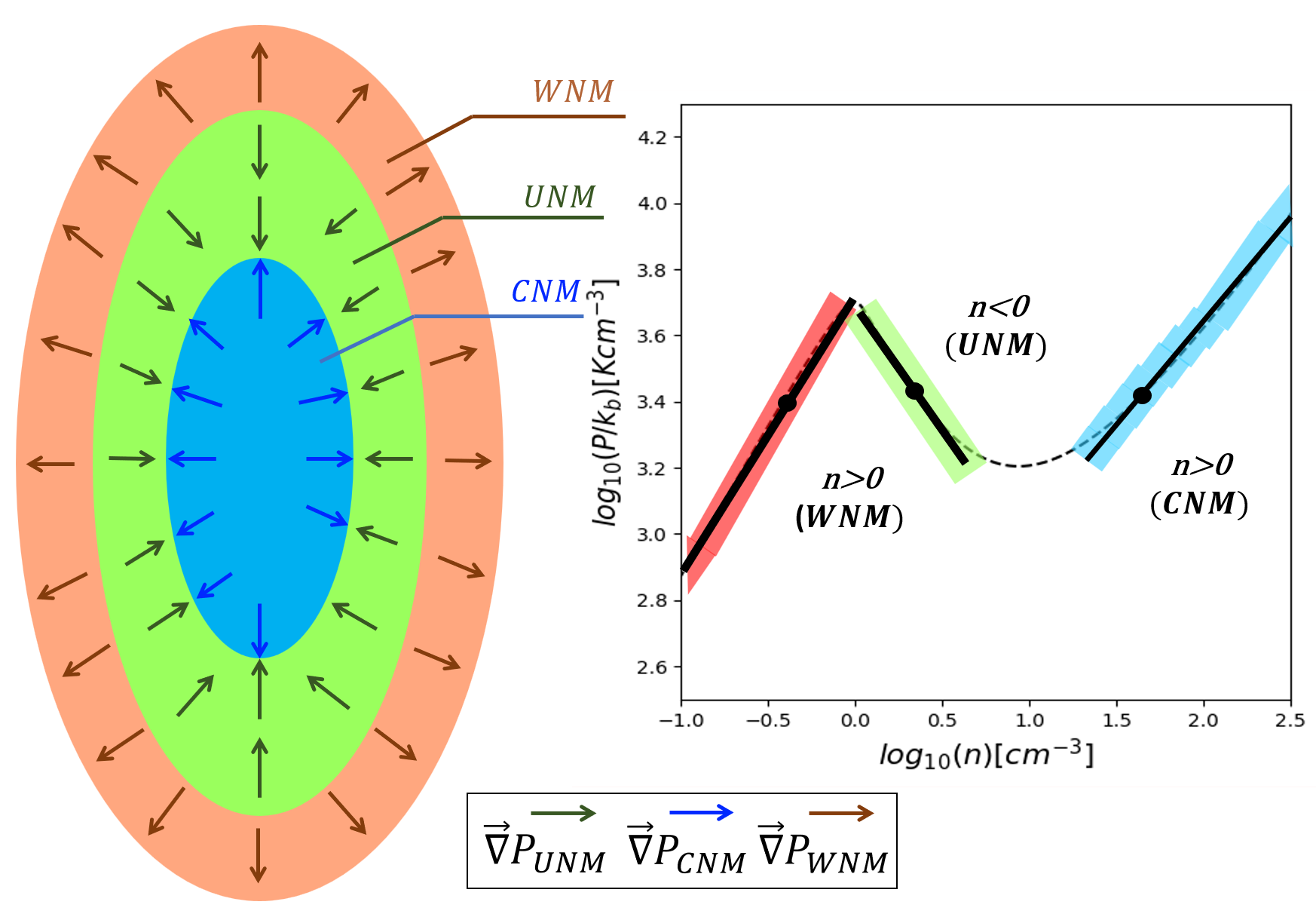}
\caption{ Left: An illustration on the direction of pressure forces for warm (orange), unstable (green) and cold (blue) phases. We can see that the pressure force direction of unstable and cold phase at the phase boundary are opposite to each other. In this paper we suggest that certain equilibrium will arrive on the phase boundary and this equilibrium will limit the aspect ratio of the filament. Right: the corresponding phase diagram illustrating how the slope of $P-\rho$ changes as we move from warm phase to cold phase. Notice that the adiabatic index for the unstable phase is negative, suggesting that its pressure force is attractive rather than repulsive. }
\label{fig:illustration}
\end{figure}

When the hydrogen number density $n_H\approx 1 cm^{-3}$ \citep{Wolfire03,Kalberla09}, the HI multiphase media enters the UNM phase until it arrives at roughly $n_H \sim 10 cm^{-3}$. In the short dynamical range, the UNM experience collapses by its own pressure term since the adiabatic index for UNM is negative. For normal EOS $P \propto \rho^{n}$, where $n > 0 $ for WNM and CNM, while $n < 0$ for UNM \footnote{\torefereeone We have to point out that the value of $n$ could be $\approx 0$ for local regions on the sky. However statistically Fig 12 of \cite{Kalberla09} suggests that $n<0$ for a wide range of values in density and galactic distance. While there could regions having $n\approx 0$, statistically $n<0$ means that there are more regions with a steeper adiabatic index. In this paper, we do not exclude any possibility that  $n<0$, but simply raise a question on {\it how} would the physics of CNM be changed when $n<0$. }. Considering the pressure force term induced by pressure gradient:
$$
\nabla P \propto n \rho^{n-1} \nabla \rho 
$$
where the sign of $n$ controls the direction of pressure force. In the case of UNM, the effect of the UNM pressure term is similar to that of the gravity term until it arrives the CNM conversion density which is roughly $10 cm^{-3}$. 

The importance of the collapsing pressure term is crucial in explaining the aspect ratio of CNM. We can have a thought experiment and consider a slab of UNM with roughly uniform density {\it by assuming the CNM forms at an abnormally unphysical high density} (No.1 of Fig.\ref{fig:cartoon}). In this scenario, the pressure force of UNM, which is analogous to the gravitational collapsing term, triggering contractions until a supporting force appears. The situation of UNM collapse is very similar to the case of Jeans instability in isothermal molecular cloud, where the balances between gravity and thermal pressure term decides the maximum size of molecular cloud stable from fragmentation. In the case of UNM, if there is no formation of CNM, the UNM will be fragmented into smaller pieces due to collapsing pressure term until the magnetic flux counterbalance the pressure contraction at smaller scale. We named this effect "UNM instability" and quantify this scale in the next subsection (\S \ref{subsec:scale}). {\torefereeone We will also call the filament formed under the force balance model as "FBM filament" to distinguish what is observed.}

However, in reality the CNM formation occurs very shorter after a short contraction of UNM. The CNM's pressure force is repulsive rather than attractive, meaning that after a short contraction the UNM's pressure force is counterbalanced by the CNM's repulsive force itself (No.2,3 of Fig.\ref{fig:cartoon}). Depending on the geometry of the CNM and the magnetic field direction. This force balance will eventually arrive. Notice that this argument does not constrain the length or the width of the resultant CNM filament, but the ratio between them. The reason behind is because when one has more CNM accumulated, the isotropic pressure force tend to decrease the aspect ratio of the filament. In the next section, we discuss analytically how the force balances between the two phases will give us the constraint on the aspect ratio of the CNM filament.

\subsection{The requirement for the polytropic index n to overcome the supporting  B-field pressure in UNM}
\label{subsec:scale}

Aside from the formation of CNM, the magnetic field also acts against the pressure collapse from UNM's EOS. It is natural to consider {\it in what condition} will the magnetic field be capable to repeal the collapse due to instability. Consider a spherical gas distribution of UNM with conservation of magnetic flux\footnote{which is adequately accurate since ambipolar diffusion does not occur in this scale} and mass in mind. In this case, we would have
\begin{equation}
\begin{aligned}
\rho = \rho_0 (\frac{r}{r_0})^{3}
\\
B   = B_0 (\frac{r}{r_0})^{2}
\end{aligned}
\end{equation}
Where $B_0,\rho_0,r_0$ are the initial magnetic field strength, density and radius of the UNM sphere respectively. We note that the introduction of the conservation of flux and mass will naturally imply $B \propto \rho ^{2/3}$. From here, we consider the pressure force term $\partial P/\partial r$ for both thermal and magnetic fields:
\begin{equation}
\begin{aligned}
\frac{\partial P_B}{\partial r} &= \frac{B^2}{2\pi r} 
\\
\frac{\partial P_{UNM}}{\partial r} &= -\frac{3n}{r} \rho_0 c^2_s (\frac{\rho_0}{\rho})^{-n}
\end{aligned}
\end{equation}
For the collapse to happen at $r=r_0$, we need $\partial P_B/\partial r <|\partial P_{UNM}/\partial r|$ i.e.
\begin{equation}
\begin{aligned}
|n| > \frac{B^2_0}{6 \pi \rho_0 c^2_s} = \frac{2}{3} (v_A/c_s)^2 
\end{aligned}
\label{eq:n}
\end{equation}
where the Alfvenic speed is given by $v_A = B^2/4\pi\rho$. We can see that Eq.\ref{eq:n} is quite sensitive to the initial B-field strength and also the temperature. To have some rough estimation of $|n|$, consider typical ISM parameters, $n_{H,0} = 1 cm^{-3}$, $B_0 = 1 \mu G$ and $T_0 = 5500 K$, we would have $|n| > 0.064$ for it to collapse. To achieve this value of $|n|$, one simply needs to stay within  {\torefereeone $R=18kpc$} from the galactic center \citep{Kalberla09}. This indicate that the collapse of USM would happen for almost all heating and cooling conditions (See, e.g. \cite{Wolfire03}).

\subsection{Instability analysis on the phase boundary: What is the maximum length of a cold neutral filament?}
\label{subsec:fb}
\begin{figure*}
\label{fig:cartoon}
\includegraphics[width=0.95\textwidth]{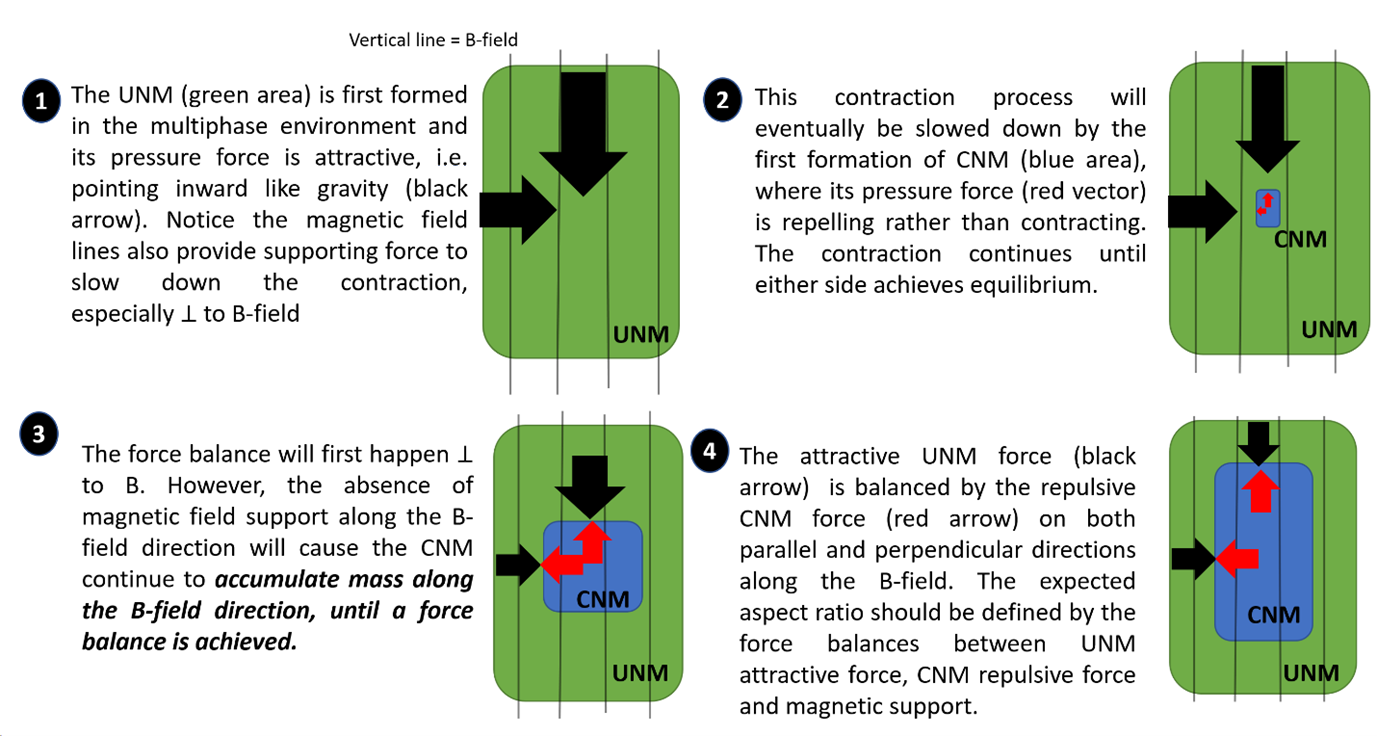}
\caption{An infographic describing the force balance model that decides the length of CNM in magnetized multiphase system.}
\end{figure*}

In this subsection, we extend our general formalism of pressure balance (\S \ref{subsec:scale}) to the case when we have a well-defined mean field along the newly formed CNM. The investigation of cloud stability has been done by \citeauthor{1991MNRAS.250..617E,1992ApJ...392..106E}( \citeyear{1991MNRAS.250..617E,1992ApJ...392..106E}, see also \citealt{2012MNRAS.423.3638I,2014ApJ...784..115I}) and later in multi-dimension by \cite{TimWaters} by considering perturbations along equilibrium. Our work considers the {\bf importance of magnetic field in 3D realistic multiphase simulations} on top of the instability analysis they developed. In this scenario, we assume that certain force equilibrium has been arrived either along or perpendicular to the mean magnetic field direction, as illustrated in panel 4 of Fig.\ref{fig:cartoon}. This equilibrium must exist and is a stable equilibrium for the following reasons:
\begin{enumerate}
    \item {\bf Existence}: Suppose not, then either UNM collapses indefinitely to form CNM, or CNM expands indefinitely. This will imply either of the fractions to be zero. From numerical simulations, we see a stable fraction of UNM during evolution (see, e.g. \cite{Kritsuk17}, see Appendix). Thus a contradiction.
    \item {\bf Stable equilibrium}: Suppose we arrive this equilibrium and we perform a slight force perturbation \footnote{Notice that the instabiltiy is considered as early as the 90's by the complexity theorists. A recent work by \cite{TimWaters} considers the coalescence instability in a numerical work and they showed that the instability persists even if all the plasma is thermally stable. They also pointed out that the instability does not require magnetic field. }. If we increase the amount of UNM, the inward force will increase the formation of CNM, causing the outward pressure force from CNM to form. Then the inward pressure force from UNM decreases, while CNM pressure force increases, which a new equilibrium will arrive. Vice versa.
\end{enumerate}

The introduction of magnetic field makes the equilibrium analysis more complicated. In MHD momentum equation, the force term of magnetic field $\Vec{F}_B$ can be written as $(\nabla \times \Vec{B}) \times \Vec{B}$. We take the z-direction as the direction parallel to B-field. In sub-Alfvenic regime, $\Vec{B}$ can be express as $\Vec{B} = b_x \hat{x} + b_y \hat{y} + B_0 \hat{z}$, where $B_0$ and $b_i \ll B_0$ are the mean field strength and random component caused by turbulence respectively. Keeping only the first order term, the magnetic field perturbation would become
\begin{equation}
\Vec{F}_B \approx B_0\frac{\partial}{\partial z}(b_x \hat{x} + b_y \hat{y})
\label{eq:Bperb}
\end{equation}

We have argued that the dense CNM exist and is stable for a long time scale, i.e. it is in a stable equilibrium. Here, we assume that the boundary of density filamentary structure is in equilibrium between thermal pressure "force" and magnetic field force. So, taking the $Dv/Dt =0$, we would have 
\begin{equation}
    \centering
    \nabla P = (\nabla \times \Vec{B}) \times \Vec{B}
    \label{eq:force_balance}
\end{equation}

One circumstance of achieving force balance would at the boundary between CNM and UNM.  In this case, a CNM filament could form and the balance would be achieved for both parallel and perpendicular direction of the CNM filament through the force balances between the CNM's repulsive force $\nabla P_C$, UNM's attractive force $\nabla P_U$ and $\vec{F}_B$. It is worth mentioning that the force balance may not be achieved along the direction of magnetic field under our treatment, since we set up Eq.\ref{eq:Bperb} under the assumption of dominant Alfven wave, i.e. no parallel-to-B perturbation components. This assumption is mildly correct since regardless of $M_s$ as we expect a dominant fraction of Alfven components \citep{CL03}. Yet it is worth mentioning that CNM generally is supersonic and sub-/trans-Alfvenic, meaning that the compressibility is rather high. The extra compression only {\bf decrease} the aspect ratio of CNM since it's much easier to compress along the B-field than perpendicularly, decreasing the originally lengthy CNM filament.  However, due to the inheritance of turbulence in UNM from the WNM, the aspect ratio for UNM is governed by the \cite{GS95} scaling. Our current analysis simply implies a tighter bound to the GS95 scaling due to the nontrivial force balances.

Notice also that the magnetic force $\Vec{F}_B$ can be through as a frictional force in this force balance model. The reason is that B-field force acts against changes, no matter whether the CNM repulsive force or UNM collapsing force is larger. That means if $|\nabla P_C| > |\nabla P_U|$, magnetic field would help to stabilize the expansion of CNM, and vice versa for $|\nabla P_U| > |\nabla P_C| $.

Notice that due to the attractive nature of the pressure force term in unstable phase, the "constant energy cascade rate" condition is invalid\footnote{Despite observations suggest that the cascade is still ongoing, see Yuen et.al (in prep)}. It is therefore difficult to analyze Eq.\ref{eq:force_balance} unless we perform an approximation here. We estimate $\nabla P \approx \Delta P / \Delta L$ and the same treatment applies to $\Vec{F}_B$. Considering the perpendicular side of filament, we would have 
\begin{equation}
\frac{\Delta P }{\Delta L} = B_0 \frac{\delta b}{l_\parallel}
\label{eq:force_balance_approx}
\end{equation}
where $l_\parallel$ denotes the parallel length scale of the filament. 

The expressions for pressure force term are different for UNM and CNM. Using the ideal gas law $ P = n_Hk_BT $, we have $\Delta P = \Delta n_H k_b T + n_H k_b \Delta T$. For CNM, we assume that is roughly isothermal\footnote{This approximation is only relative, since in our formalism $\delta T$ for UNM is much larger than that of CNM. }, which makes  $\Delta P = \Delta n_H k_b T_C = (n_{peak} - n_C) k_b T_C$. Here $n_{peak}, n_C$, and $T_C$ denote the peak HI number density for the filament, minimum CNM formation density and average CNM temperature respectively. For UNM, $\Delta P_U $ can be approximated as :
\begin{equation}
\Delta P_U= (n_C - n_W) k_b T_C + n_C k_b (T_C-T_W)
\end{equation}
Here, $n_W,T_W $ represent the density and temperature of WNM. For simplicity, we take $n_C = 10 n_W$ and $T_W = 27.5T_C$. That implies $n_C = 10 cm^{-3}$ and  $T_C = 200K$ when $n_W = 1 cm^{-3}$ (Wolfire et.al 2003) and  $T_W = 5500K$ (Kalberla et.al 2009) in terms of cgs unit. After that, $\Delta P_U$ could be simplified as
\begin{equation}
\Delta P_U \approx  -25.6 n_c k_b T_C
\end{equation}
where the minus sign here indicates its attractive force nature. Notice that $\Delta L$ varies when we move from the unstable phase to the cold phase. For instance, along the $\perp$ direction of CNM, $\Delta L_C = l_{\bot}/2$ if we calculate the gradient at the center of CNM. We also define $\Delta L_U = l_{U,\bot}$ as the thickness of ambient UNM layer. It is trivial to see that the amplitude of pressure force depends on the thickness of both CNM filament and its ambient UNM layer.

In below, we consider a situation where a CNM filament is expanding until arriving an equilibrium. In this scenario, its expanding force together with magnetic field force could be balanced by the UNM's attractive force. We call this the UNM compression regime. Fig.\ref{fig:cartoon} shows an cartoon to illustrate the whole process. The force balance expression would then be
\begin{equation}
\frac{25.6 n_c k_b T_C}{l_{U,\bot}} \approx \frac{2(n_{peak}-n_c)k_b T_C}{l_{\bot}} +  B_0 \frac{\delta b}{l_\parallel}
\end{equation}

The aspect ratio  $l_\parallel/l_\bot$ can be easily derived by rearranging the equation:
\begin{equation}
\frac{l_\parallel}{l_\bot} \mu m_H n_c c^2_s ( 25.6  \frac{l_\bot}{l_{U,\bot}} - 2(\frac{n_{peak}}{n_C}-1))= B_0 \delta b
\end{equation}
where $\mu m_H c^2_s = k_B T_C$ with $c_s$ denotes the sound speed of the CNM. Although we do not have the value of $\delta b$, the Alfvenic Mach Number $M_A \approx \delta b / B_0$ controls the relative strength of the magnetic field perturbation compared to the mean field. In addition, the Alfvenic speed $V_A$ could be written as $V^2_A=B_0^2/n_C \mu m_H $ in the cgs unit. The expression of aspect ratio would then become
\begin{equation}
\frac{l_\parallel}{l_\bot} = \frac{v_A^2}{c^2_s} M_A [ 25.6  \frac{l_\bot}{l_{U,\bot}} - 2(\frac{n_{peak}}{n_C}-1) ]^{-1}
\end{equation}
Below we use the expressions of sonic $M_S = V^2/c^2_s$ and Alfvenic Mach number $M_A = V/V_A$ here. The final expression for CNM aspect ratio would become:
\begin{equation}
\frac{l_\parallel}{l_\bot} = \frac{M_S^2}{M_A} [ 25.6  \frac{l_\bot}{l_{U,\bot}} - 2(\frac{n_{peak}}{n_C}-1) ]^{-1}
\label{eq:aspect_UNM_compression}
\end{equation}

\begin{figure}
\label{fig:model}
\includegraphics[width=0.5\textwidth]{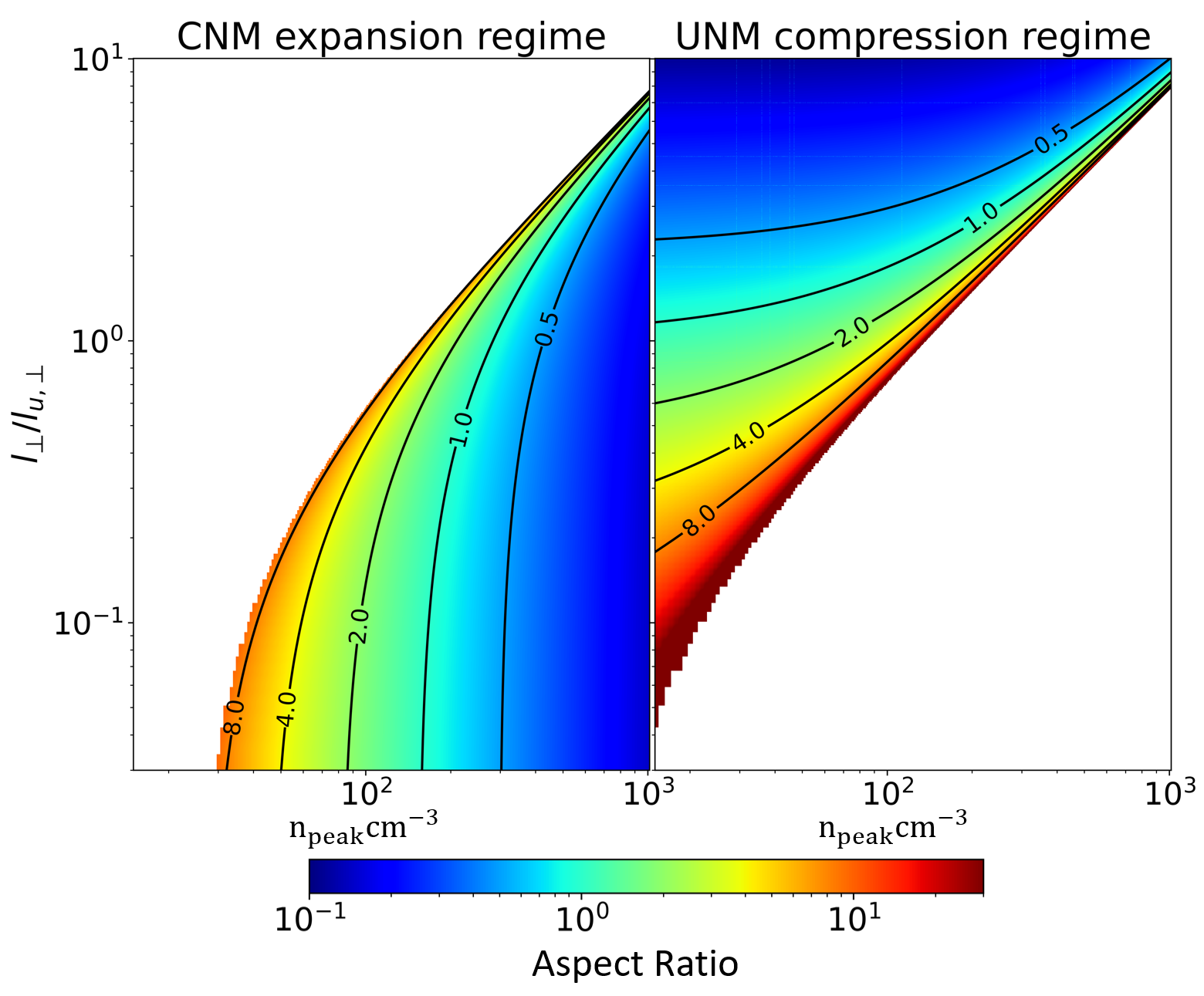}
\caption{The analytical prediction (Eq.\ref{eq:aspect_CNM_expansion}) of how the aspect ratio varies as functions of peak hydrogen number densities of the cold and unstable phases ($n_{peak}$), and also the ratio of widths of CNM and UNM ${l_\bot}/l_{U,\bot}$ using the typical Mach numbers for CNM $M_S\approx5$,$M_A\approx0.8$. The iso-contours denoting the values of the aspect ratio. When the aspect ratio $>1$ the filament is parallel to B, while $<1$ means the filament is perpendicular to B. }
\end{figure}

\subsection{The interpretation of the force balancing model}

Eq.\ref{eq:aspect_UNM_compression} shows that there are a few important factors that impacts the expected aspect ratio of CNM. First of all, the ratio $M_s^2/M_A$ defines the base aspect ratio of the filament. The $1/M_A$ dependence is similar to that of what \cite{Xu19} found in their paper, but this factor appeared here for a completely different reason: it is the force balance rather than the cascade constancy in defining the length of the CNM filament. There are some extra factors in our model: (1) the ratio between the peak and average cold neutral media density $n_{peak}/n_C$, and (2) the ratio between the width of the filament over the width of the ambient UNM shell  $l_\bot/l_{U,\bot}$ further restrain the aspect ratio of each individual CNM filament. The former parameter is decided by the maximum density where the HI-H2 conversion occurs. For CNM this parameter is roughly 10-30. The second parameter is decided by the dynamics of the gases. In our work, we did not find any constraints on the parameter $l_\bot/l_{U,\bot}$. Therefore we shall treat it as a free parameter. For a stable envelope-shell structure (Fig.\ref{fig:illustration}), the three sets of parameters ($M_s^2/M_A$,$n_{peak}/n_C$,$l_\bot/l_{U,\bot}$) decides the length of the CNM filament. There are two regimes in terms of the force balance, namely the UNM compression regime and the CNM expansion regime, which is defined by the relative magnitudes of the pressure forces:

{\noindent \bf UNM compression regime} For this regime,  $|\nabla P_U|>|\nabla P_C|$, the aspect ratio is given by Eq.\ref{eq:aspect_CNM_expansion} (left) and Eq.\ref{eq:aspect_UNM_compression} (right), which restrict the domain of  $l_\bot/l_{U,\bot}$ to be 
\begin{equation}
0 < \frac{2}{25.6}(\frac{n_{peak}}{n_c}-1)<\frac{l_\bot}{l_{U,\bot}}.
\end{equation}

{\noindent \bf CNM expansion regime}
Otherwise, we would arrive in another regime, CNM expansion regime, which means $|\nabla P_C|>|\nabla P_U|$, which the expression of aspect ratio would become
\begin{equation}
\frac{l_\parallel}{l_\bot} = \frac{M_S^2}{M_A} [ 2(\frac{n_{peak}}{n_C}-1) - 25.6  \frac{l_\bot}{l_{U,\bot}} ]^{-1}.
\label{eq:aspect_CNM_expansion}
\end{equation}

Also, the domain will be restricted in 

\begin{equation}
0 < \frac{l_\bot}{l_{U,\bot}}<\frac{2}{25.6}(\frac{n_{peak}}{n_c}-1).
\end{equation}

To obtain $l_\bot/l_{U,\bot} < 1 $, we have a maximum CNM mean peak density $n_{peak,0} < 150 cm^{-3}$.

Fig. \ref{fig:model} shows an prediction of aspect ratio of the two regimes  (Eq.\ref{eq:aspect_CNM_expansion} and Eq.\ref{eq:aspect_UNM_compression}) with varying $l_\bot/l_{U,\bot}$ and $n_{peak}$ at $M_S\approx5$, $M_A\approx 0.8$ which is the typical conditions for CNM. Notice that the GS95 criterion allows the filament\footnote{GS95 (Sub-Alfvenic):$\frac{l_\parallel}{l_\perp}\sim M_A^{-4/3}\left(\frac{L}{l_\perp}\right)^{1/3} $} to have a mean length of 13 following Xu et.al (2019), assuming $(L/l_\perp)^{1/3} \sim 10$ (See Appendix).  One can see from the figure, the model tends to produce an filament with relative short aspect ratio ($2 \sim 8$). More importantly, the mean aspect ratio (5.3, See \S 3) is significantly shorter than what the GS95 predicted (13). In fact, only a narrow parameter space will allow the formation of CNM filament with long aspect ratio.  This indicates the CNM filament would be clumpy rather than long filamentary shape in the 3D space.

\section{Numerical results in 3D}
\label{sec:num}
\subsection{Multi-phase simulation setup}
\begin{table*}[]
\label{table:simulation} 
\begin{tabular}{@{}llllllll@{}}
\toprule
                     & \multicolumn{3}{c}{\begin{tabular}[c]{@{}c@{}}Weak field case\\ $n_{0} = 3 m_h cm^{-3}$\\ $B_0 = 1 \mu G$\end{tabular}} & \multicolumn{1}{c}{} & \multicolumn{3}{c}{\begin{tabular}[c]{@{}c@{}}Strong field Case\\ $n_{0} = 5 m_h cm^{-3}$\\ $B_0 = 5 \mu G$\end{tabular}} \\ \midrule
                     & WNM                            & UNM                            & CNM                             &                      & WNM                             & UNM                             & CNM                             \\
$M_s$                   & 0.7                            & 2.7                            & 8.1                               &                      & 0.57                            & 2.1                             & 5.0                             \\
$M_A$                   & 0.8                            & 1.6                            & 2.2                             &                      & 0.09                            & 0.28                            & 0.74                           \\
Mass filling factor $M_{C/U/W}$ & 0.08                           & 0.28                           & 0.63                            &                      & 0.05                            & 0.33                            & 0.62                            \\
Volume filling fator $V_{C/U/W}$ & 0.61                           & 0.31                           & 0.070                           &                      & 0.54                            & 0.38                            & 0.075                           \\ \bottomrule
\end{tabular}
\caption{Simulation parameters and the statistics of each phase. The resolution for the strong field simulation is $512^3$, and that for the weak field case is $480^3$.}
\end{table*}

To compare the result in section \ref{sec:theory}, we use the 3D MHD multi-phase simulations generated from the MHD code Athena++, which were also being used in \cite{Yuen21a} and Ho et al. in prep. For the initial state, we set up a 3D periodic turbulence box with the length of 200 pc and we are assuming the fluid represents the bulk neutral hydrogen in the interstellar media. We adopt the realistic cooling and heating function proposed by \cite{KI02}. {\torefereeone The simulation was originally constant in density and was driven via spectral velocity perturbation in the Fourier space. At around 110Myr the turbulence box has produced a realistic multiphase medium, which the parameters are listed in Tab.\ref{table:simulation}. {\torefereetwo The phase diagram of "strong" case has been shown at Fig.\ref{fig:phase_diagram}.} 
We set up two simulations with the conditions similar to the realistic multi-phase neutral hydrogen gases with mass/volume fractions being consistent with observations. We call two simulations used in this paper as the "strong field case" and the "weak field case". The strong field case have a higher mean number density of $n_H = 5 cm^{-3}$ and stronger magnetic field strength $B_0 = 5\mu G$, while that for the weak field case is  $n_H = 3cm^{-3}, B_0 = 1\mu G$. We shall define the gas as the cold phase when the temperature of the gas is below 200K while those above 5500K as to warm phase, while the gas in between is the unstable phase. Table \ref{table:simulation} shows the detail parameters and statistics of each simulation.

\begin{figure}
\label{fig:phase_diagram}
\includegraphics[width=0.47\textwidth]{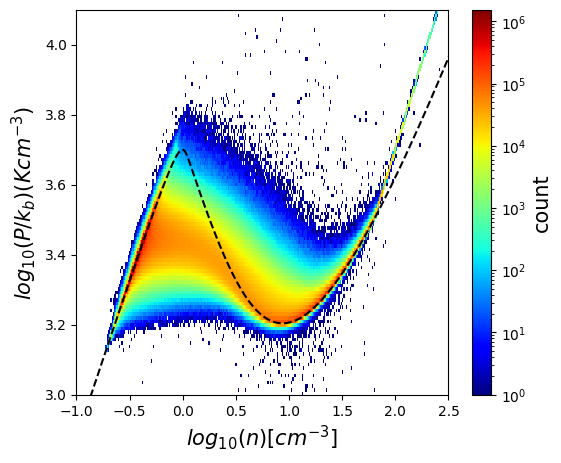}
\caption{phase diagram of our strong B-field case}
\end{figure}

To quantify the aspect ratio of CNM {\torefereetwo dense structure}, we set up both temperature and density thresholds to filter out the low density, non-cold structure. After that, we apply a periodic labelling algorithm to detect isolated individual structures. We define the sub-structure detected with more than 500 pixels in volume {\torefereetwo as the candidate of filamentary structure. \footnote{We note that the choice of 500 pixel may not be fully enough to resolve the highly elongated density structure as those filamentary structure's pixel count may not be enough to be captured by our current choice. This is the weigh between the statistical sampling size of the dense structure and resolution of the simulation. As the simulation are only $512^3$ and the CNM is often fragmented and occupy few percent of the space, the number of dense structure with high pixel count are limited. a future study with higher resolution simulation could improve the aspect ratio measurement.}}
We further define the maximum density value of each filament as $n_{peak}$. To determine the aspect ratio of a filamentary structure, we define a $3\times3$ inertia matrix P \footnote{Notice that the matrix $P = IR^2-I_{moment}$, where $R^2 = \sum_{i} x_ix_i+y_iy_i+z_iz_i$ while $I_{moment}$ is the density-free moment of inertial matrix used in classical mechanics. The physical interpretation of this P matrix is simply to characterize the principle axis of the filaments. } as:

$$
P = \begin{bmatrix}
I_{xx} & I_{xy} & I_{xz}\\
I_{yx} & I_{yy} & I_{yz}\\
I_{zx} & I_{zy} & I_{zz}
\end{bmatrix},
$$
where $I_{xy} = \sum_i x_iy_i$ with $x_i,y_i,z_i$ to be the coordinates of individual pixels in filamentary structure {\it relative to the center of mass of the filament}. The matrix P has a set of eigenvalues $\lambda_i$ and we define the aspect ratio R to be $R = max(\lambda)/min(\lambda)$ if all $\lambda$ are positive. 

\subsection{Multi-phase structure and CNM filament axis-ratio in simulation}
Throughout the evolution of both simulations, we observed the same findings from previous numerical studies\citep{Kritsuk17}. With the presence of magnetized turbulence, we notice that a significant amount of UNM continues to exist until the end of simulation, in terms of both mass and volume fraction, which is shown in table \ref{table:simulation} (See also Appendix).

We further observe the 3D density structure at the simulation. The morphology of dense structure exist in the form of filamentary shape and mostly elongated along the mean field direction. However, such a "filament" is actually not entirely in one phase. The UNM phase resides at the outer shell of the filament with a higher aspect ratio, while CNM lies in core of the filament but with a shorter aspect ratio. The extracted filament's structure is onion-like, as illustrated at Fig. \ref{fig:illustration}: With warm and diffuse WNM shell further outside and the converging UNM form as a intermediate layer, a cold and dense CNM sit as a stable core at the center of the filament. On the left of Fig. \ref{fig:Simulation_Plot} shows an example of such picture but in real MHD multi-phase simulation. We zoom into a selected region (middle of Fig.\ref{fig:Simulation_Plot}) where we see several CNM formed within the UNM. Notice that the CNM on the top left corner of the middle of Fig.\ref{fig:Simulation_Plot} is not parallel to the mean field, but it is {\torefereeone misaligned}. In the right panel of  Fig. \ref{fig:Simulation_Plot}, pressure gradients of UNM converge (the black vector within the green area) into the embedded CNM core (blue area) with expanding pressure gradient. In addition, we observe that the ratio between the thickness of CNM core and UNM shell, $l_\bot/l_{U,\bot}$, is also not a constant in our simulations.

For strong field case, we pick $T=200K$ and $n_H = 30 cm^{-3}$ as the threshold of the filament filtering process. The numbers are justified since in our simulations the filaments are entering the cold phase at {\torefereeone $T\sim 184 K$} and $n_H\sim 10 cm^{-3}$. The choice of a higher threshold is to make the filaments disconnected. A total of 9664 sub-structures are found through this method but only about 472 candidates has the size large than 500 pixels (equivalent to a volume of 36 $pc^3$). From all the candidates, the mean of the HI peak density $\langle n_{peak} \rangle$ is 114 $cm^{-3}$ and the numerical mean of aspect ratio from all filamentary are 5.32. 

We note that it is possible to determine the mean aspect ratio through our model by substituting $\langle n_{peak} \rangle$, $\langle l_\bot/l_{U,\bot}\rangle$, $M_{S,C}$ and $M_{A,C}$ into equation Eq.\ref{eq:aspect_CNM_expansion}. We can roughly estimate $\langle l_\bot/l_{U,\bot}\rangle$ as $((1+V_U/V_C)^{-1/3}-1)^{-1}$ using the values of the volume fractions of CNM $V_C$ and UNM $V_U$ from Table \ref{table:simulation}\footnote{We assume the gas is a two layer gas sphere, with CNM sits at the centre with radius $l_{\bot}$ and UNM as a outer layer with radius $l_{\bot}+l_{\bot,U}$. Notice that $V_U+V_C \propto (l_{\bot}+l_{\bot,U})^3$ and $V_C\propto l_{\bot}^3$. By knowing the volume fraction of the inner and outer region, we can get $l_{\bot}/l_{\bot,U}$.}, which yields the prediction of mean aspect ratio to 3.77 with discrepancy of $28.9\%$ comparing to measured value 5.32. However, we note that the histogram of the aspect ratio is actually left-skewed, with the peak very close to our {\torefereeone theoretical} prediction (3.47). 

To explore further the statistical distribution of the aspect ratio as a function of the three parameters ($M_s^2/M_A$,$n_{peak}/n_C$,$l_\bot/l_{U,\bot}$), we plot the $n_{peak}-R$ scatter plot with both of their histograms as Fig.\ref{fig:Simulation_Plot} for our strong field model (See Table \ref{table:simulation}). For the given values of the CNM turbulence parameter ($M_s \approx 5, M_A \approx 0.8$), we color the main panel as shaded grey assuming a range of possible values of $l_\perp/l_{U,\perp} \in [0,5,3.0]$ and the given $n_{peak}/n_C$ for these filaments.  We found out $445/472$ data points falls within grey area, which suggest most of the data point could be explained by our UNM force balancing model. To compare our results with the GS95 one, in the Appendix we use an isothermal simulation of similar turbulence condition and extraction parameter and we found that the average length of the filaments in the isothermal case is twice of that in the CNM case. Both of these evidences indicate that the GS95 is not a proper model in explaining the aspect ratio of the CNM filaments. The analysis of the super-Alfvenic regime is included in the Appendix.

\subsection{How does our result compared to the isothermal counterpart? }

\begin{figure*}
\label{fig:GS95_isothermal}
\includegraphics[width=0.47\textwidth]{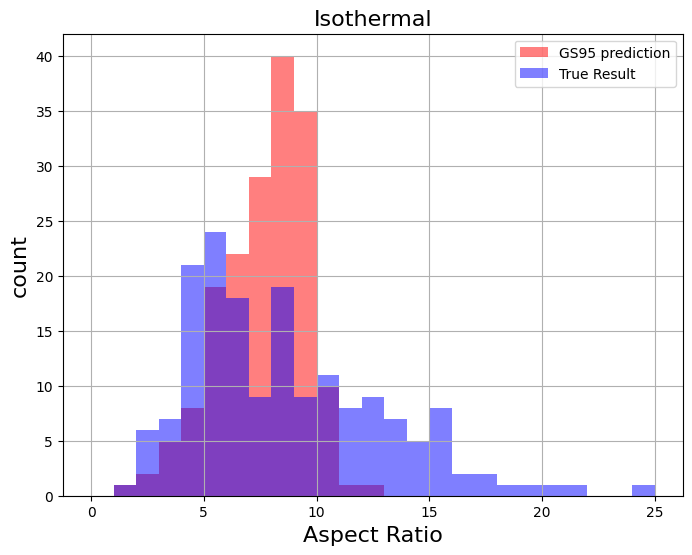}
\includegraphics[width=0.47\textwidth]{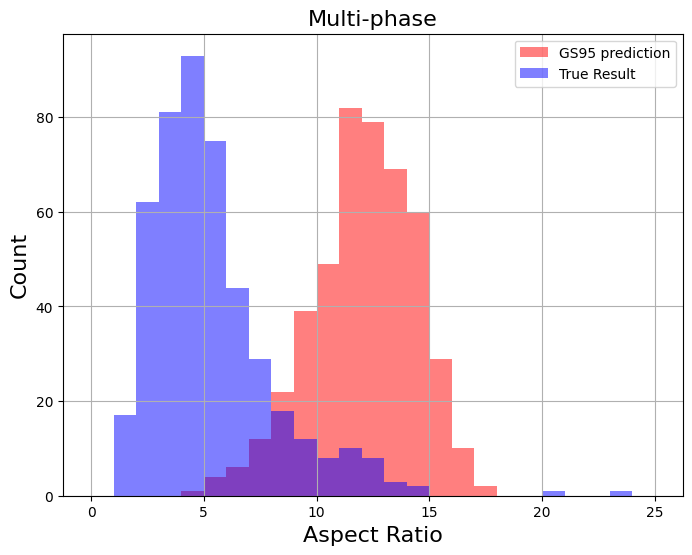}
\caption{histogram of aspect ration between true filament aspect ratio and GS95 prediction ($\frac{l_\parallel}{l_\perp} \sim M_A^{-4/3} (\frac{L_{inj}}{l_\perp})^{1/3}$). \\
Left: isothermal case ($M_S \approx 6, M_A \approx 0.8, N^3 = 792^3$. Right: Multi-phase strong {\torefereeone B-field} case}.
\end{figure*}

In the previous section, we stress that the aspect ratio results for the CNM in our multiphase simulations are shorter than GS95 prediction. A prominent question is, do the aspect ratio of filaments in the isothermal case follows GS95, and how does that different from the CNM with the same turbulence condition? To answer this question, we prepare an isothermal simulation with similar turbulence conditions as the CNM and compare the aspect ratio between the isothermal and multiphase simulations.  Fig \ref{fig:Model} shows the distribution of aspect ratio for filaments in multiphase (left) and isothermal case (right). Notice that in isothermal simulations there are no explicit threshold in determining a filament, as opposed to CNM where there is a clear cutoff in phase diagram. However, due to the self-similarity of MHD turbulence, we can extract the over-density from isothermal simulations as long as the filaments are statistically significant. We can see that for the isothermal case the mean value of the filament aspect ratio roughly follows the expectations of GS95. This is very different from the case of multiphase simulations, where the mean value of the filament aspect ratio is only 1/2 of that of GS95 prediction. This simple comparison confirms our argument in the paper that it is not GS95 that decides the length of the CNM filament.

Notice that , the dispersion of the filament aspect ratio in our isothermal numerical simulations are roughly twice of GS95 expected. We do not have the reason why this is the case. Some of the possible reasons are : (1) The filament extraction method might extract shock-related filaments who naturally do not follow GS95. In this case these high compression shocks contributes to the high end of the aspect ratio histogram. (2) We filtered structures that are volumetrically small, which these tiny features are likely to have smaller aspect ratio. Yet, the deviation of the GS95 prediction and the numerics in our numerical simulations do not change our conclusion that the multiphase counterpart is much more deviated from GS95. 

\begin{figure*}
\includegraphics[width=0.8\textwidth]{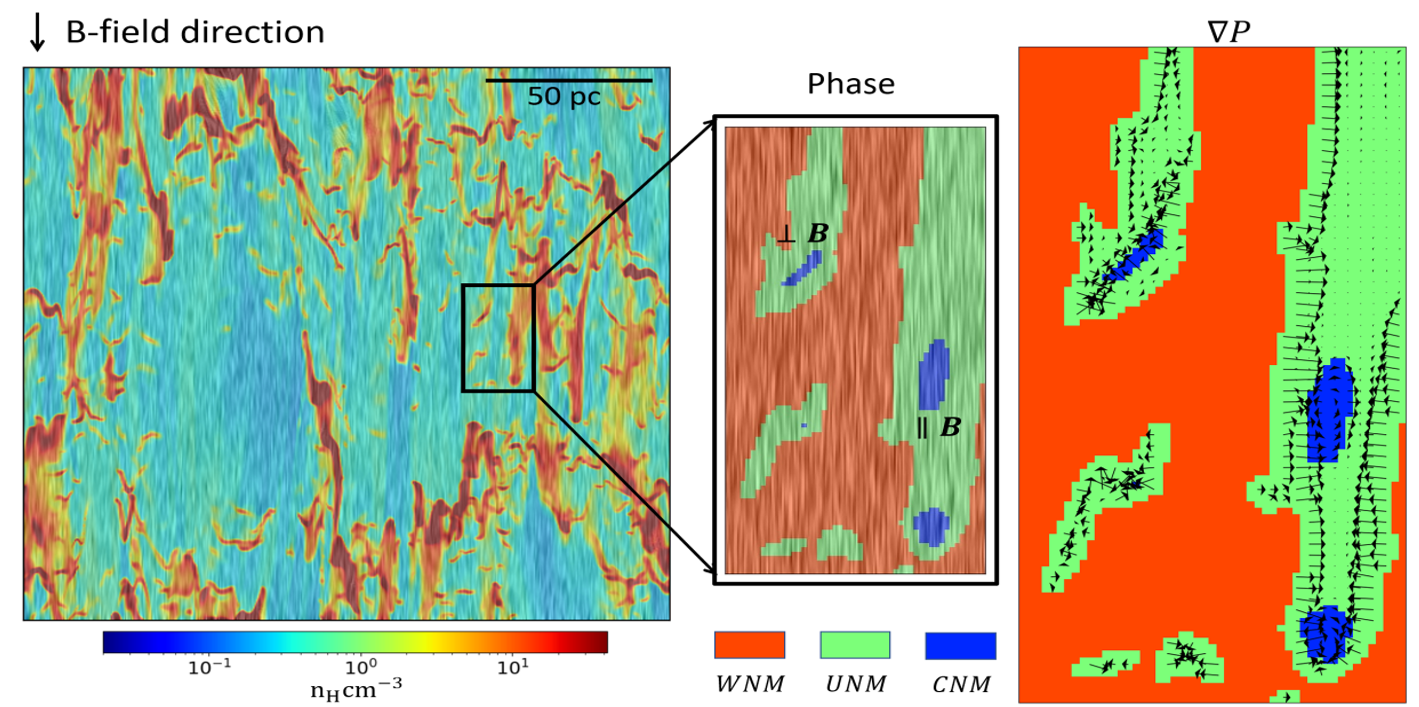}
\caption{Left: 3D density plot in a MHD multiphase simulation overlaid with magnetic field. Middle: Phase recognition diagram from a selected local region based on the cut-off from the right panel of Fig 2 with magnetic field directions. This region features two different types of the CNM: the upper left has a perpendicular CNM filament formed due to collapses of UNM; and on the right-hand side a parallel CNM filament is formed. One can very easily recognize that the elongated features are UNM (green), but not CNM (blue region). CNM is generally more roundish. Right: Pressure gradient (UNM/CNM) overlaid with phase recognition diagram. The pressure force for UNM is pointing towards the CNM, while that of CNM is repelling, the balances of these two forces determine the geometry of CNM.}
\label{fig:Simulation_Plot}
\end{figure*}

\begin{figure*}
\label{fig:Model}
\includegraphics[width=0.8\textwidth]{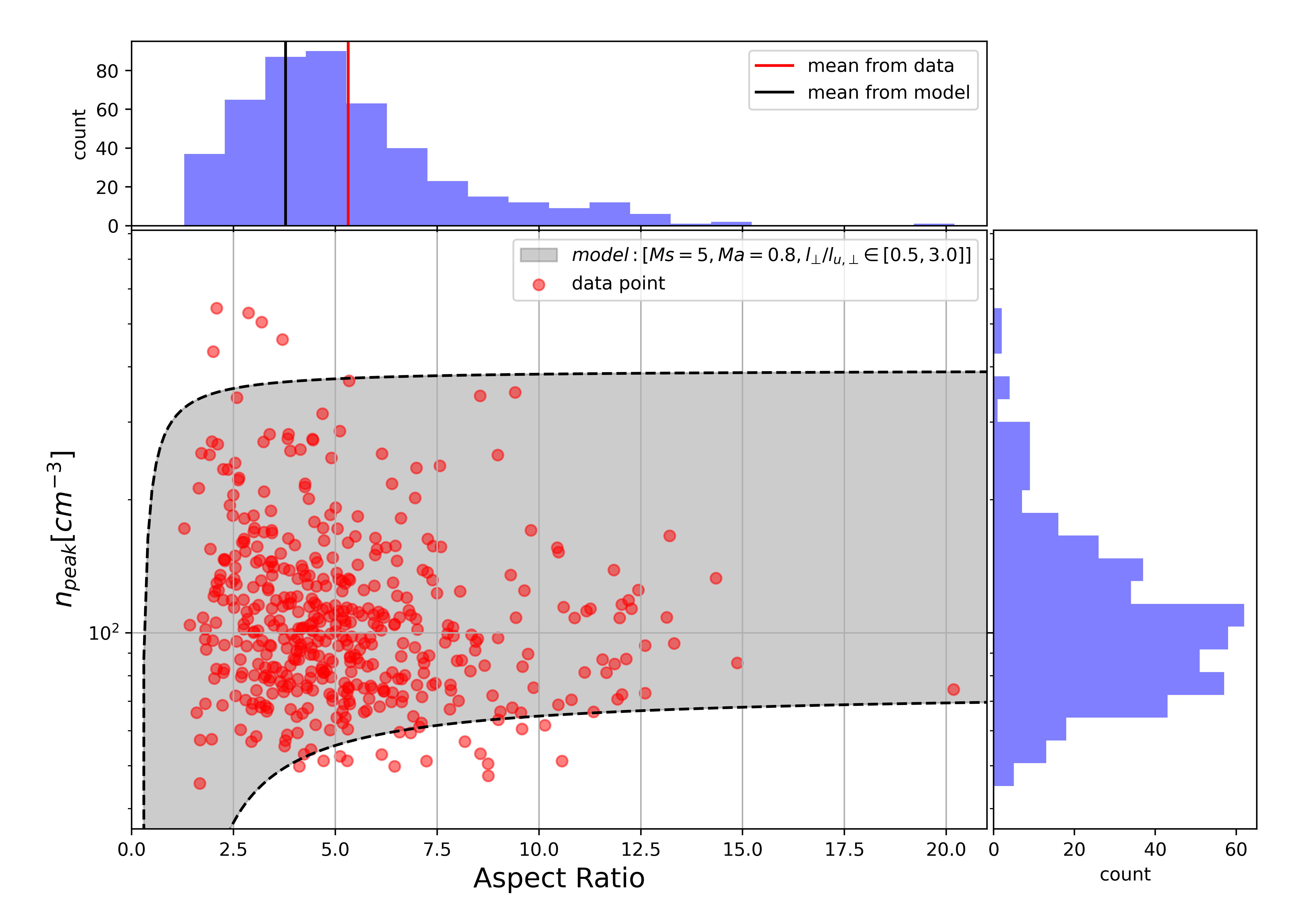}
\caption{A figure showing the scatter plot (red points) of peak number density towards the aspect ratio of the filament with both of the histograms plotted in sub-panels. The grey area indicates the prediction of force balance model for a range of possible values of $l_\perp/l_{U,\perp} \in [0,5,3.0]$.}
\end{figure*}

\section{Discussions}
\label{sec:dis}
\subsection{Observational diagnostics of CNM filament aspect ratio and applications}
The core result of this paper suggests that the aspect ratio of CNM filament is subjected to the force balance at the boundary of UNM and CNM. This leads to our conclusion that in typical turbulent environments the aspect ratio of CNM filament has a lower value (See Eq. \ref{eq:aspect_CNM_expansion}, Appendix) than GS95. This value is rarely larger than 20 (Fig.\ref{fig:Model}). This aspect ratio is inconsistent with the recent findings of HI filaments in observations which has an aspect ratio of more than 100. The question is, why there is an inconsistency between our theory and observations in terms of the aspect ratio of HI filaments? For the simplicity of our discussion, we would only discuss the properties of parallel filaments and defer the perpendicular one into future papers (Ho et.al in prep)) since we do not usually see HI filaments that are perpendicular to B-field very often in observations.


We argue that the parallel filaments that are seen in observations could be a mixture of both UNM and CNM, similar to what happened in Fig.\ref{fig:Simulation_Plot}. The reason is, while there is a upper limit of the aspect ratio for CNM under our force balance model (Eq\ref{eq:aspect_CNM_expansion}), there is no such restriction for UNM. In fact, as we noticed from the theory section, the aspect ratio induced by GS95 cascade is roughly twice of that is given by the force balance model. Even with the inclusion of the GS95 model, the CNM itself under its natural turbulent condition would not have an aspect ratio of 100. In this scenario, one of the most possible scenario here is that the observationally "identified" CNMs are actually UNM features with CNM to be core. Notice that this postulate is entirely compatible to the absorption study from \citep{PC2019} in terms of the location of the CNM, because the CNM is embedded within the UNM anyhow (See Fig.\ref{fig:Simulation_Plot} for an illustration). Moreover, recent absorption study \citep{Murray18} suggests that the lower temperature part of UNM also contributes significantly to the absorption lines. If we imagine a scenario where the CNM is actually fully self-absorped, the leftover ambient shell of UNM will be then very long and thin on the sky, which could be one of the explanation of why we see such features everywhere in HI maps. However, these features will then not fulfill the conditions of being actually cold in terms of the phase diagram. 

In fact, our analysis in this paper further explains why the formed molecular clouds rarely have that high aspect ratio until the arrival of Jeans criterion of facing strong compression. If there is a very long CNM formed on the sky violating the conditions that we derived in this paper, either the GS95 criterion or our UNM instability criterion will naturally destroy these ong filaments.


Notice that right after the formation of CNM, the conversion from HI to H2 starts to happen. This conversion is rapid \citep{BS16} and happens at a density threshold not really far away from the CNM formation density. This means that some of the higher density fluctuations of HI have been converted into H2 in this scenario. This does not impact the aspect ratio permitted for CNM under our force balancing model.

\subsection{The importance of MHD turbulence in shaping the CNM filaments}

GS95 provides an expression for the aspect ratio of filaments formed from isothermal, magnetized turbulence. It has been long postulated that the same cascade applies to multiphase HI but no evidence is provided until recently (Yuen et.al in prep). Yet, from this paper we see that the aspect ratio of CNM filaments are far lower than the expectation from GS95. 

Two of the very important assumptions that GS95 required are the constancy of the cascade rate and the maintenance of critical balances. Notice that the existence of UNM negative force term implies that the energy cascade rate is not constant, similar to the case of gravity as illustrated from Li et.al (2018). While UNM occupies only a short range of wavenumber only in terms of spectrum (Yuen et.al in prep), that means the CNM filaments formed should subject to the criteria aside of GS95.

\subsection{\torefereeone Our model compared to CNM formation model based on thermal instability}
{\torefereeone One of the leading model attempting to explain the formation of CNM is the thermal instability (TI) model based on the anisotropic conduction (See, e.g. Choi \& Stone 2012). We would like to compare our model to theirs for completeness. One of the important aspect from the TI model is that the natural introduction of anisotropic conduction term: $\kappa \propto \hat{B}\hat{B}\cdot \nabla T$ will lead to elongated features in simulations (See, e.g. Fig 8 of Choi \& Stone 2012). This is a popular model in explaining the huge and cold filaments seen in intra-cluster media studies. However, we have to note that (1) turbulence conduction (due to diffusion) is actually way stronger than that of thermal conduction in the case of ISM (See Cho et al. 2003). In fact, if we plug in the physical parameters at the stage of UNM or CNM, the diffusion coefficient from turbulence is decades in order larger than that of thermal conduction. (2) The anisotropic conduction does not introduce a universal mechanism in "chopping the cold filaments into segments" in the case of weak magnetization. Despite CNM in the studies from Choi \& Stone shows segmentation in the case of strong field simulation, for their weak field simulations the cold filaments are extending like waves. One of the very important observational aspect for CNM is: they are actually in segments, but not infinitely long wave-like structures dangling on the sky. Both GS95 models (Xu et.al 2019) and our model provide a mechanism for the cold filaments in weak magnetization to fragment. In particular, our model explicitly suggests that filaments longer than the criterion outlined in the current paper must fragment. We believe that the reason why the CNM in the presence of strong B-fields in Choi \& Stone's work is due to our UNM dynamical instability model. (3) The consideration of continuous driving is crucial for the understanding the importance of UNM instability. Without continuous driving, the multiphase turbulence will quickly collapse into two-phases with minimal (mass/volume) fractions for UNM. Yet, the existence of continuous injection allows one to maintain a stable fraction of UNM, and as a result the UNM dynamical instability that we outlined in this paper continues to be valid over the course of CNM formation (See Appendix \ref{app:fraction}). This is also not considered in TI models as the driving of turbulence will immediately make the TI conduction be overshadowed by turbulence conduction (See Cho et al. 2003).

}

\subsection{Caveat of our work and potential improvements}

In our paper, the force balance between CNM, UNM and magnetic field is the essential element to model the aspect ratio (See Eq.\ref{eq:aspect_CNM_expansion},Eq.\ref{eq:aspect_UNM_compression}). However,  the force balance condition is not an accurate description when turbulence is present since the latter also provides support forces to the filamentary feature. Yet as we know, turbulence support is similar to magnetic field support in this scenario, which acts like a friction term in slowing down the collapse of UNM. Our force balance model could include the turbulence support rather easily through similar fashion as the magnetic field force. Yet, the filamentary structure could be away from equilibrium. Our model does not include such consideration and aspect ratio could diverge from the prediction. Yet as we can see from Fig.\ref{fig:Model} our model bounds most of the filaments within our expectation curve.  Furthermore, although we know a stable amount of UNM exists in the ISM system, there is a lack of theoretical study to understand the relationship between the turbulence and UNM, such as the energy transfer and the energy cascade rate in UNM. Due to the prolonged lifetime of UNM, a detailed study of the properties of UNM with the presence of magnetized turbulence is urgently need in the near future (Ho et.al in prep)

In addition, while most of filamentary structure lie within the prediction range of the model, only $472$ out of 9963 sampled disconnected features are qualified as filament, since majority of isolated CNM structures are less then 500 pixels in volume from our simulation. In our current study, we discard these very tiny isolated structures due to the presence of numerical dissipation. However notice that both GS95 and our model predicts that the filament aspect ratio tends to be longer when we are going to smaller scale. Our simulations can trace filaments with thickness up to $1 pc$  but simulation with higher resolutions might provide better insight whether the filaments at the thickness of $0.1pc$ are actually very filamentary.

{\torefereeone We also acknowledge that in realistic ISM the turbulence is driven by supernovae explosion. In terms of the theory of MHD turbulence, this is equivalent to consider a compressible driving (Walch et al. 2015, Kim \& Ostriker 2018, Seifried et al. 2020, Rathjen et.al. 2021). We note that the input driving only changes the fraction of the relative MHD modes in the system. The statistics of MHD turbulence due to the change of relative fraction of modes have been studied by a number of authors in the community (See, e.g. Cho \& Lazarian 2003, Federrath et.al 2015).  However even for incompressible driving, there is a considerable amount of compressible modes present, and for our case the CNM is likely falling into this regime as the sonic Mach number is large. Yet a study on comparing the formation probability of CNM in the case of different driving mechanism is a worthwhile study.}

\section{Conclusion}
\label{sec:con}

In this paper, we utilize the force balance model in explaining why CNM should have a much lower aspect ratio than what observationally suggested. We discover a new instability triggered by the attractive nature of the UNM pressure term, and recognize a force balance along the UNM-CNM phase boundary. To summarize,
\begin{enumerate}
    \item The consideration of attractive nature UMM pressure term is necessary to understand who CNM has a given aspect ratio. (S\ref{sec:theory})
    \item From our instability analysis, the mean value of the aspect ratio of the CNM is given by Eq \ref{eq:aspect_CNM_expansion}, and it is significantly smaller than the GS95 prediction (Appendix)
    \item Our result indicates that the observationally seen filaments are either having way stronger magnetic field conditions than we previously expected, or those are simply not CNM but more likely to be the UNM ambient shell with CNM self-absorbed. 
\end{enumerate}

{\noindent \bf Acknowledgment} KHY acknowledges Chris Mckee for his inspirational comments on the nature of filamentary CNM aspect ratio. We also acknowledge Snezana Stanimirovic and Tim Waters for their valuable comments. {\torefereeone We thank Hui Li for providing extensive comments on the physics of cold filaments on the sky.} KWH, KHY \& AL acknowledge the support the NSF grant AST 1212096 and NASA grant NNX14AJ53G. The main simulations and the first version of the work is done during KHY's tenure in UW Madison. {\torefereeone Research presented in this article was supported by the Laboratory Directed Research and Development program of Los Alamos National Laboratory under project number(s) 20220700PRD1.}\linebreak

{\noindent \bf Data Availability} 
The data underlying this article will be shared on reasonable request to the corresponding author.

\appendix
\section{Appendix}
\label{sec:Ap}

\subsection{Fraction of the phase across evolution time}
\label{app:fraction}

\begin{figure*}
\label{fig:UNM_Stat}
\includegraphics[width=0.8\textwidth]{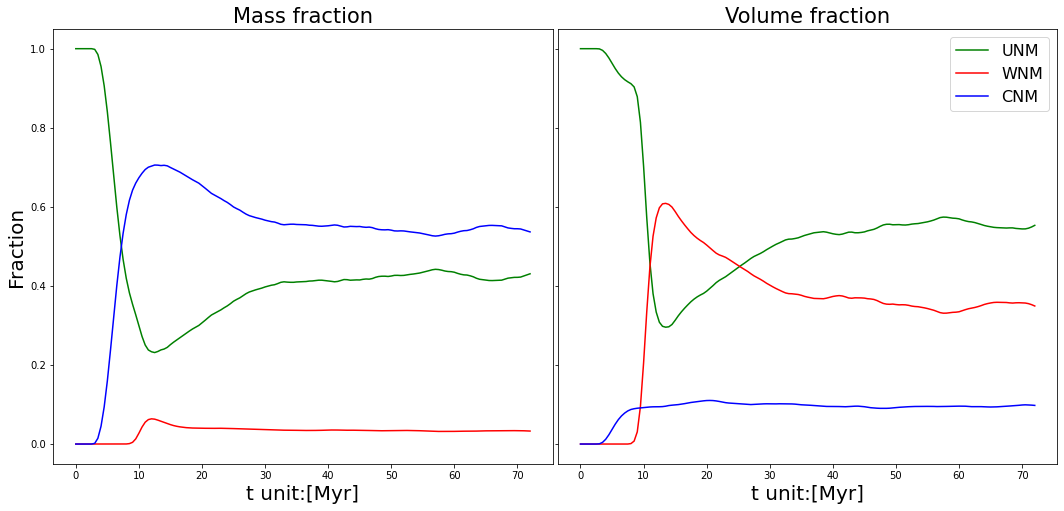}
\caption{A figure showing volume fraction and mass fraction of each phase across the simulation time described in Appendix \ref{app:fraction}.}
\end{figure*}
In the main text, we stress that the UNM will have a stable mass and volume fraction across the simulation time if turbulence is introduced \citep{Kritsuk17}. To demonstrate this simple fact, we {\torefereeone dump out the time series of the "strong" case of our multi-phase simulations (See Tab.\ref{table:simulation})}. We set the density and B-field strength of the new simulation same as the strong field case . Fig. \ref{fig:UNM_Stat} shows the variation of the mass and volume fraction of each phase across the simulation time. One can see that, with the presence of MHD turbulence, the ISM system reaches an equilibrium within 30 Myr. After that, the fraction of each phases maintain in a steady amount until the end of the simulation. Notice that the natural UNM cooling rate is less than 1Myr without the presence of turbulence, suggesting the crucial role of turbulence in supporting the large UNM fraction as seen in observations \citep{Kalberla18}.

\subsection{The aspect-ratio of CNM filaments in super-Alfvenic case}

In this section, We extend our analysis from strong field case (mean field = $ 5 \mu G$) to weak field case (mean field = $ 1\mu G$). The parameters of the simulation are stated in Table. \ref{table:simulation}. To extract the filament, we follow the method used in the main text. We pick $T = 200K$ and $n_H = 20 m_H cm^{-3}$ as the threshold of the filament filtering process. The mean of the peak core density $n_{peak}$ is $70 m_H cm^{-3} $. Fig. \ref{fig:Model_WeakB} shows the $n_{peak}-R$ scatter plot. The mean aspect ratio for the extracted filamentary structures in the weak field numerical simulations are 5.96. Although most of the individual data points, which represents the aspect ratio of each filament, lie within the model prediction (the grey area of Fig.\ref{fig:Model_WeakB}, the model's prediction for the mean value is $1.75$, which has a discrepancy of $73.6\%$ compare to actual value.   

Comparing to the strong field case, we notice that the model's mean prediction deviate significantly from the actual average value. This is not surprisingly to us because in the main text (\S \ref{sec:theory}) we utilized the fact that $\delta B/B \ll 1 $ across the CNM to construct the model. In this case, the whole environment, including both the CNM and its ambient transient UNM, should be in sub-Alfvenic regime, i.e. the Alfvenic mach number for CNM $M_{A,CNM}$ for strong field case (See Table \ref{table:simulation}) $< 1$. The reason we limited our study to sub-Alfvenic case in the main text is that, CNM is believed to be highly magnetized and sub-Alfvenic, which is well interpreted from different observational papers (e.g. see \citealt{Clark15}).

However, for our the weak field case $M_{A,CNM}$ is higher than one ($M_{A,CNM}\sim 2.2$) due to the increase of densities during transitions. The analysis would then fail due to the violation of the requirement $M_{A,CNM}<1$. Yet, the usual magnetic field strength in ISM is much stronger than our weak field case ($\sim 1\mu G$), and the $M_A$ would less than one in the case but we recognize that the super-Alfvenic regime possibly exists. So, we shall refer our study to later publications.

\begin{figure*}
\label{fig:Model_WeakB}
\includegraphics[width=0.8\textwidth]{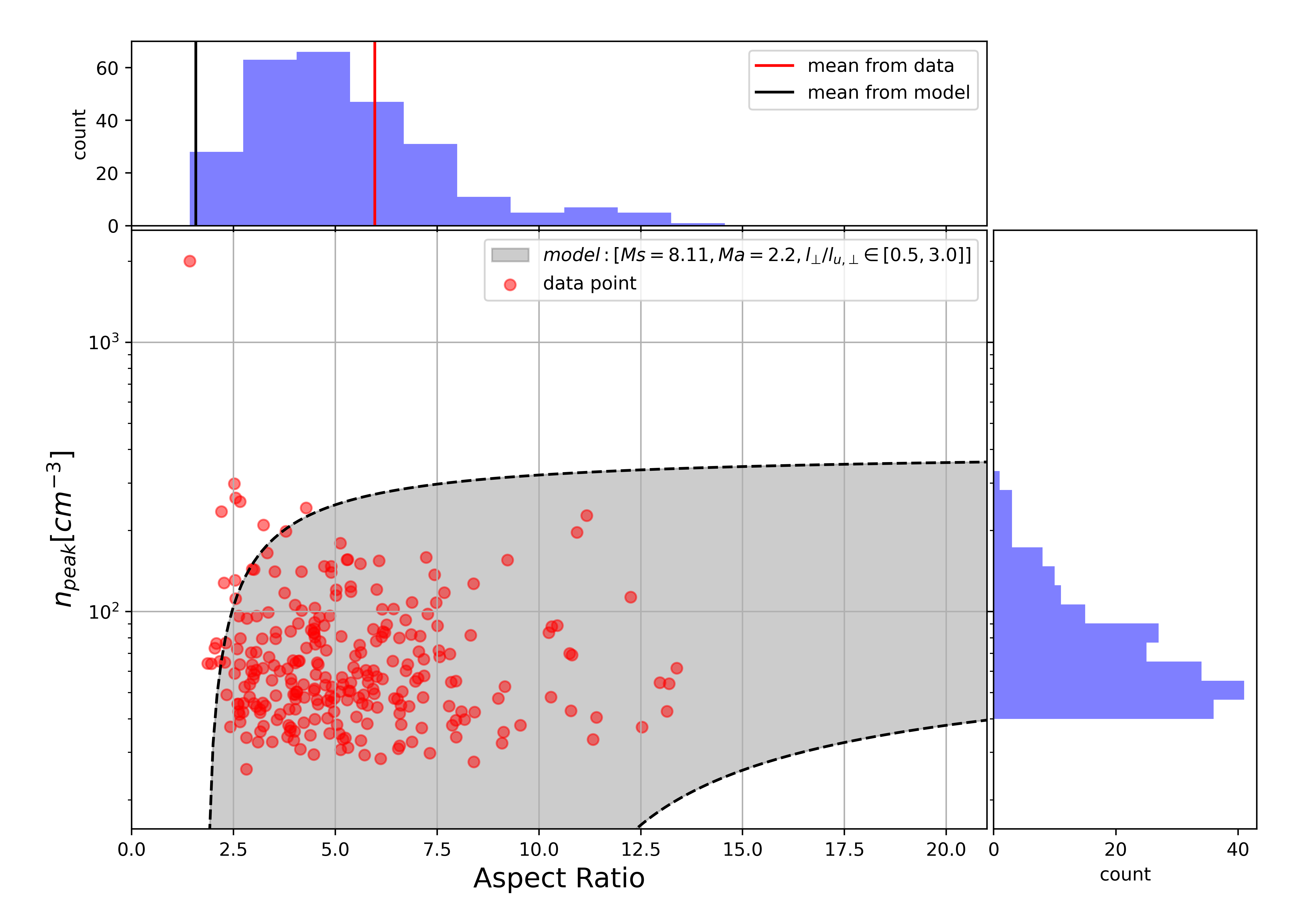}
\caption{A figure showing the scatter plot (red points) of peak number density towards the aspect ratio of the filament with both of the histograms plotted in sub-panels. The grey area indicates the prediction of force balance model for a range of possible values of $l_\perp/l_{U,\perp} \in [0,5,3.0]$.}
\end{figure*}



\bsp	
\label{lastpage}
\end{document}